\documentclass{svproc}
\usepackage{graphicx}%
\usepackage{multirow}%
\usepackage{amsmath,amssymb,amsfonts}%
\usepackage{mathrsfs}%
\usepackage[title]{appendix}%
\usepackage{xcolor}%
\usepackage{textcomp}%
\usepackage{manyfoot}%
\usepackage{booktabs}%
\usepackage{algorithm}%
\usepackage{algorithmicx}%
\usepackage{algpseudocode}%
\usepackage{listings}%
\usepackage{tikz}
\usepackage{url}

\def\orcidID#1{\unskip$^{[#1]}$} 

\begin{document}
\mainmatter              
\title{Cooperation guides evolution in a minimal model of biological evolution.}
    \titlerunning{Cooperation guides evolution in a minimal model of biological evolution.}  
%
\author{Conor Houghton
\orcidID{0000-0001-5017-9473}}
\authorrunning{Conor Houghton} 

\institute{Intelligent Systems Laboratory, University of Bristol, Michael Ventris Building,
Bristol, BS8 1UB, United Kingdom.\\
\email{conor.houghton@bristol.ac.uk},\\ WWW home page:
\texttt{http://conorhoughton.github.io}}

\maketitle              

\begin{abstract} A challenging simulation of evolutionary dynamics based on a three-state cellular automaton is used as a test of how cooperation can drive the evolution of complex traits. 
Building on the approach of Wolfram (2025), the fitness of a cellular automaton is determined by the number of updates it endures before returning to an empty configuration. 
In our population-based model, groups of automata are assessed either by summing individual fitness or by the maximum fitness in the group; this second scheme mimics cooperation by allowing fit rules to support their less fit group mates. 
We show that cooperation substantially improves the population's ability to evolve longer-lived rules, illustrating a Baldwin Effect for Cooperation: higher diversity across individuals unlock otherwise inaccessible evolutionary paths. 
\end{abstract}
\section{Introduction}\label{sec1}
This paper uses a simple but demanding test of evolutionary dynamics to support the proposal made in \cite{Houghton2024} that cooperation can guide evolution. Cooperation, according to this proposal, can bridge otherwise difficult evolutionary gaps. Very much like the Baldwin Effect  \cite{Baldwin1896,Morgan1896,Waddington1942}, this mechanism supports the discovery and acquisition of complex multi-gene characteristics that standard evolutionary processes alone would attain only very slowly. Thus, the proposal does not suppose that information is transferred back into the genetic code through cooperation: it does not depart from standard Darwinian evolution in which the genetic code of a specific organism is not changed by the organism's behaviour or the events that it experiences. However, the proposal does not simply assert the established truth that cooperation can be of benefit, \cite{AxelrodHamilton1981,Nowak2006,WestGriffinGardner2007}, it describes an additional benefit of cooperation: cooperation leads to an evolutionary dynamics better able to discover and acquire complex traits.

The Baldwin Effect refers to the phenomenon by which learned behaviors can facilitate and accelerate evolutionary change. In the version emphasized in \cite{HintonNowlan1987}, learning effectively ``broadens'' the evolutionary path to complex, otherwise improbable traits: individuals can temporarily acquire beneficial characteristics through learning, so even partial genetic configurations can be viable, thereby increasing the likelihood that the fully adaptive gene set is eventually found and fixed by natural selection. In his original paper, \cite{Baldwin1896}, Baldwin himself also highlighted a second, complementary aspect: because learned abilities enable more organisms to survive and reproduce even in absence of a complete genetic basis for their advantageous abilities, the overall diversity within a population remains higher. This, in turn, promotes a broader and more efficient evolutionary search. Over generations, traits that were initially learned can become genetically stabilized -- a process sometimes termed genetic assimilation -- ultimately intertwining learned and inherited factors in the shaping of complex adaptations \cite{Waddington1942,Waddington1953,Crispo2007}.

In what will be referred to as the Baldwin Effect for Cooperation, the value of individual traits can accumulate in a group, giving value to traits which may have no value in isolation. It may be helpful to consider a just-so cartoon example. Picture a flock of jungle fowl occasionally threatened by snakes. One mutation helps a bird recognize a snake but, by itself, that trait accomplishes nothing if the bird cannot effectively defend itself; a different mutation makes the bird instinctively stand tall to chase off snakes, but this ability is useless if the bird never detects the threat. In the absence of cooperation these two traits are hard to acquire since they so rarely occur together. However, if one hen squawks a warning while another already knows how to scare snakes, the group cooperates to fend off danger even without every hen having both mutations. Cooperation in this case has two benefits for evolution, first, it adds value to each of the two traits which in isolation would not be as useful and second, since not all the animals need both traits, the flock can maintain a more varied genetic profile, allowing a broader exploration of the evolutionary landscape.

In \cite{Houghton2024} the Baldwin Effect for Cooperation was demonstrated using a very simple example, one based on the treatment of the original Baldwin Effect presented in \cite{HintonNowlan1987}. A similar proposal for a Baldwin Effect for Symbiosis was given in \cite{WatsonPollack1999}; this also provided a simple demonstration. The purpose of this paper is to test the effect in a richer and more challenging simulation of evolutionary dynamics, one in which the second potential benefit --- greater diversity in genetic profile --- can be observed. The simulation is based on the recent proposal that one-dimensional cellular automata can be used to study evolution \cite{MitchellCrutchfieldHraber1994,Wolfram2002,Wolfram2025}.

Specifically, a three-state cellular automaton is used. This is a discrete dynamical system where each cell can take one of three possible values, often labeled 0, 1, or 2, with the 0-state regarded as `empty'. At each time step, a simple update rule, applied to every cell in parallel, determines the cell's new state based on its current state and the states of its two nearest neighbors. An example of a rule is given in Fig.~\ref{fig:rule}. This example is intended to illustrate both the operation of the model and one of its remarkable properties. It is clear that for a specific starting configuration some rules will `carry on forever': in this paper, as in \cite{Wolfram2025}, the starting configuration always has a single cell in the 1-state and all other cells empty. With red representing the 1-state and blue the 2-state, an obvious example of a rule that carries on forever is
\begin{center}
    \begin{tikzpicture}[
  x=3.25mm,
  y=3.25mm,
  cell/.style={
    draw=black,
    thick,
    minimum size=3.25mm,
    inner sep=0,
    outer sep=0
  }
]

\def\StateColor#1{%
  \ifnum#1=0 white\else%
  \ifnum#1=1 red\else%
    blue\fi\fi
}
    \node[cell, fill=\StateColor{0}]   at (0,0) {};
    \node[cell, fill=\StateColor{1}]    at (1,0) {};
    \node[cell, fill=\StateColor{0}]  at (2,0) {};

    \node[cell, fill=\StateColor{1}] at (1,-1) {};
\end{tikzpicture}

\end{center}
with all other clauses giving an empty state. In contrast, the rule whose only non-trivial clause is
\begin{center}

\begin{tikzpicture}[
  x=3.25mm,
  y=3.25mm,
  cell/.style={
    draw=black,
    thick,
    minimum size=3.25mm,
    inner sep=0,
    outer sep=0
  }
]

\def\StateColor#1{%
  \ifnum#1=0 white\else%
  \ifnum#1=1 red\else%
    blue\fi\fi
}
    \node[cell, fill=\StateColor{0}]   at (0,0) {};
    \node[cell, fill=\StateColor{1}]    at (1,0) {};
    \node[cell, fill=\StateColor{0}]  at (2,0) {};

    \node[cell, fill=\StateColor{2}] at (1,-1) {};

\end{tikzpicture}

\end{center}
gives a cellular automaton with a single non-empty update. The remarkable property is that there are rules which neither carry on forever, nor stop after only a few steps: they continue for multiple steps before stopping. The rule in Fig.~\ref{fig:rule} is one such example; it has 977 updates before reaching the empty state. 

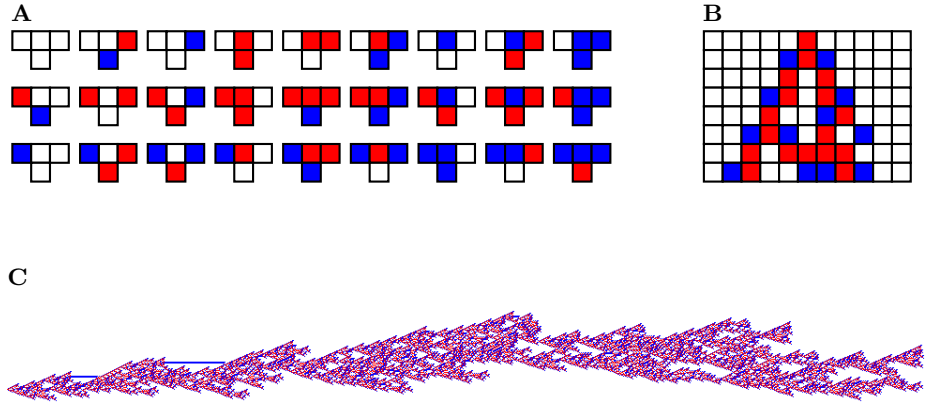
\begin{figure}[t]
\begin{tabular}{lll}
  \begin{tabular}[t]{@{}l@{}}
    \textbf{A}\\
    \begin{tikzpicture}[
  x=2.5mm,
  y=2.5mm,
  cell/.style={
    draw=black,
    thick,
    minimum size=2.5mm,
    inner sep=0,
    outer sep=0
  }
]

\def\rulearray{{0,2,0,1,0,2,0,1,2,2,0,1,1,2,2,1,1,2,0,1,1,0,2,0,2,0,1}}

\def\StateColor#1{%
  \ifnum#1=0 white\else%
  \ifnum#1=1 red\else%
  blue\fi\fi
}

\foreach \x in {0,...,8}
{
\foreach \y in {0,...,2}
{

  \pgfmathsetmacro{\i}{9*\y + \x} 
    \pgfmathsetmacro{\minusy}{2-\y}     
  \pgfmathtruncatemacro{\leftd}{\i/9}         
  \pgfmathtruncatemacro{\temp}{mod(\i,9)}
  \pgfmathtruncatemacro{\midd}{\temp/3}       
  \pgfmathtruncatemacro{\rightd}{mod(\i,3)}   

  \pgfmathtruncatemacro{\ruleval}{\rulearray[\i]}

 \begin{scope}[xshift=0.9*\x cm,yshift=0.75*\minusy cm] 

    \node[cell, fill=\StateColor{\leftd}]   at (0,0) {};
    \node[cell, fill=\StateColor{\midd}]    at (1,0) {};
    \node[cell, fill=\StateColor{\rightd}]  at (2,0) {};

    \node[cell, fill=\StateColor{\ruleval}] at (1,-1) {};


  \end{scope}
}
}

\end{tikzpicture}
  \end{tabular}
  &
  \hspace{0.9cm}
  \begin{tabular}[t]{@{}l@{}}
    \textbf{B}\\
    \begin{tikzpicture}[
  x=2.5mm,
  y=2.5mm,
  cell/.style={
    draw=black,
    thick,
    minimum size=2.5mm,
    inner sep=0,
    outer sep=0
  }
]

\def\StateColor#1{%
  \ifnum#1=0 white\else%
  \ifnum#1=1 red\else%
    blue\fi\fi
}
 \node[cell, fill=\StateColor{0}]   at (0,0) {};
    \node[cell, fill=\StateColor{0}]   at (1,0) {};
    \node[cell, fill=\StateColor{0}]    at (2,0) {};
    \node[cell, fill=\StateColor{0}]  at (3,0) {};
    \node[cell, fill=\StateColor{0}]   at (4,0) {};
    \node[cell, fill=\StateColor{1}]    at (5,0) {};
    \node[cell, fill=\StateColor{0}]  at (6,0) {};
    \node[cell, fill=\StateColor{0}]   at (7,0) {};
    \node[cell, fill=\StateColor{0}]    at (8,0) {};
    \node[cell, fill=\StateColor{0}]  at (9,0) {};
    \node[cell, fill=\StateColor{0}]  at (10,0) {};

 \node[cell, fill=\StateColor{0}]   at (0,-1) {};
    \node[cell, fill=\StateColor{0}]   at (1,-1) {};
    \node[cell, fill=\StateColor{0}]    at (2,-1) {};
    \node[cell, fill=\StateColor{0}]  at (3,-1) {};
    \node[cell, fill=\StateColor{2}]   at (4,-1) {};
    \node[cell, fill=\StateColor{1}]    at (5,-1) {};
    \node[cell, fill=\StateColor{2}]  at (6,-1) {};
    \node[cell, fill=\StateColor{0}]   at (7,-1) {};
    \node[cell, fill=\StateColor{0}]    at (8,-1) {};
    \node[cell, fill=\StateColor{0}]  at (9,-1) {};
    \node[cell, fill=\StateColor{0}]  at (10,-1) {};

 \node[cell, fill=\StateColor{0}]   at (0,-2) {};
        \node[cell, fill=\StateColor{0}]   at (1,-2) {};
    \node[cell, fill=\StateColor{0}]    at (2,-2) {};
    \node[cell, fill=\StateColor{0}]  at (3,-2) {};
    \node[cell, fill=\StateColor{1}]   at (4,-2) {};
    \node[cell, fill=\StateColor{0}]    at (5,-2) {};
    \node[cell, fill=\StateColor{1}]  at (6,-2) {};
    \node[cell, fill=\StateColor{0}]   at (7,-2) {};
    \node[cell, fill=\StateColor{0}]    at (8,-2) {};
    \node[cell, fill=\StateColor{0}]  at (9,-2) {};
    \node[cell, fill=\StateColor{0}]  at (10,-2) {};
    
 \node[cell, fill=\StateColor{0}]   at (0,-3) {};
        \node[cell, fill=\StateColor{0}]   at (1,-3) {};
    \node[cell, fill=\StateColor{0}]    at (2,-3) {};
    \node[cell, fill=\StateColor{2}]  at (3,-3) {};
    \node[cell, fill=\StateColor{1}]   at (4,-3) {};
    \node[cell, fill=\StateColor{0}]    at (5,-3) {};
    \node[cell, fill=\StateColor{1}]  at (6,-3) {};
    \node[cell, fill=\StateColor{2}]   at (7,-3) {};
    \node[cell, fill=\StateColor{0}]    at (8,-3) {};
    \node[cell, fill=\StateColor{0}]  at (9,-3) {};
    \node[cell, fill=\StateColor{0}]  at (10,-3) {};

 \node[cell, fill=\StateColor{0}]   at (0,-4) {};
        \node[cell, fill=\StateColor{0}]   at (1,-4) {};
    \node[cell, fill=\StateColor{0}]    at (2,-4) {};
    \node[cell, fill=\StateColor{1}]  at (3,-4) {};
    \node[cell, fill=\StateColor{0}]   at (4,-4) {};
    \node[cell, fill=\StateColor{0}]    at (5,-4) {};
    \node[cell, fill=\StateColor{2}]  at (6,-4) {};
    \node[cell, fill=\StateColor{1}]   at (7,-4) {};
    \node[cell, fill=\StateColor{0}]    at (8,-4) {};
    \node[cell, fill=\StateColor{0}]  at (9,-4) {};
    \node[cell, fill=\StateColor{0}]  at (10,-4) {};

 \node[cell, fill=\StateColor{0}]   at (0,-5) {};
        \node[cell, fill=\StateColor{0}]   at (1,-5) {};
    \node[cell, fill=\StateColor{2}]    at (2,-5) {};
    \node[cell, fill=\StateColor{1}]  at (3,-5) {};
    \node[cell, fill=\StateColor{2}]   at (4,-5) {};
    \node[cell, fill=\StateColor{0}]    at (5,-5) {};
    \node[cell, fill=\StateColor{1}]  at (6,-5) {};
    \node[cell, fill=\StateColor{0}]   at (7,-5) {};
    \node[cell, fill=\StateColor{2}]    at (8,-5) {};
    \node[cell, fill=\StateColor{0}]  at (9,-5) {};
        \node[cell, fill=\StateColor{0}]  at (10,-5) {};

 \node[cell, fill=\StateColor{0}]   at (0,-6) {};
    \node[cell, fill=\StateColor{0}]   at (1,-6) {};
    \node[cell, fill=\StateColor{1}]    at (2,-6) {};
    \node[cell, fill=\StateColor{0}]  at (3,-6) {};
    \node[cell, fill=\StateColor{1}]   at (4,-6) {};
    \node[cell, fill=\StateColor{1}]    at (5,-6) {};
    \node[cell, fill=\StateColor{1}]  at (6,-6) {};
    \node[cell, fill=\StateColor{1}]   at (7,-6) {};
    \node[cell, fill=\StateColor{0}]    at (8,-6) {};
    \node[cell, fill=\StateColor{0}]  at (9,-6) {};
        \node[cell, fill=\StateColor{0}]  at (10,-6) {};

 \node[cell, fill=\StateColor{0}]   at (0,-7) {};
    \node[cell, fill=\StateColor{2}]   at (1,-7) {};
    \node[cell, fill=\StateColor{1}]    at (2,-7) {};
    \node[cell, fill=\StateColor{0}]  at (3,-7) {};
    \node[cell, fill=\StateColor{0}]   at (4,-7) {};
    \node[cell, fill=\StateColor{2}]    at (5,-7) {};
    \node[cell, fill=\StateColor{2}]  at (6,-7) {};
    \node[cell, fill=\StateColor{1}]   at (7,-7) {};
    \node[cell, fill=\StateColor{2}]    at (8,-7) {};
    \node[cell, fill=\StateColor{0}]  at (9,-7) {};
        \node[cell, fill=\StateColor{0}]  at (10,-7) {};
  
\end{tikzpicture}
  \end{tabular}
\end{tabular}\\[1cm]
\noindent \textbf{C}
\begin{center}
  \includegraphics[width=\textwidth]{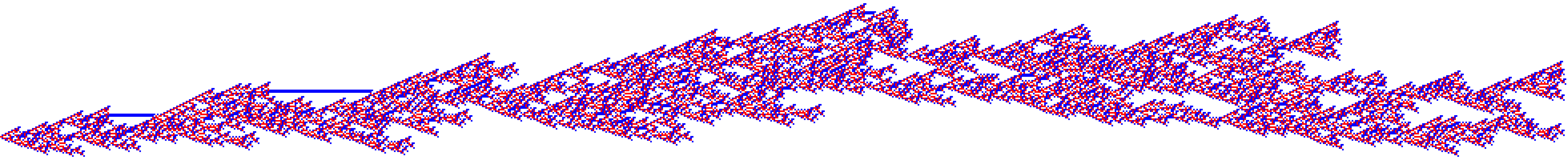} 
\end{center}
\caption{\textbf{The rule 1811860148953 cellular automaton}. The rule for this automaton is given in \textbf{A}; the sequence of new states regarded as a base-3 number gives the rule name: 1811860148953. The initial sequence of updates is given in \textbf{B} while \textbf{C} gives the entire sequence until the rule results in the empty state; for convenience this graph has been rotated anti-clockwise so `time' runs from left to right with the initial state on the left.\label{fig:rule}}
\end{figure}

In \cite{Wolfram2025} it is proposed that this system is useful for studying the dynamics of biological evolution. The idea is that the fitness of a rule is given by the number of updates, starting from a single 1-state, before it returns to the empty state, a rule that carries on forever is regarded as having a fitness of -1. The advantage of this system is that the fitness landscape is complex; some changes to the rule can increase its fitness, others decrease it, but the significance of these changes are very variable, a change in one clause of the rule can easily lead to a cellular automaton that goes on forever and therefore have -1 fitness. 

A simple version of evolution is used in \cite{Wolfram2025}. In this system the agent starts with a trivial rule in which every clause maps to the empty state. At every generation a single randomly chosen clause is mutated to give a different outcome. If this change reduces the fitness, this change is discarded and the agent reverts to its previous rule. If it improves, or, crucially, preserves the fitness, the mutation is retained. Retaining neutral changes allows for evolutionary exploration. The very start of evolution gives a simple example for understanding these dynamics. For the empty rule there is no single change which can increase fitness, every mutation either results in a rule which carries on forever, or immediately terminates in the empty state. Thus, if evolution relied on only retaining beneficial changes, the rule would remain static. However, there are many mutations that do not change the fitness, this is common at this early stage because many clauses are never used in a world with only one 1-state. As these neutral changes are accepted, they `build-up' allowing evolution to eventually discover a non-trivial rule. 

In this paper this evolutionary scheme is extended to schemes in which the Baldwin Effect for Cooperation can be assessed. In practice this involves a population of agents, each with its own rule and fitness. The agents are placed in groups, the least fit group is culled at each generation and a new group is spawned by breeding and mutating the agents in the most fit group. There are two schemes used for these dynamics, in the non-cooperative scheme the fitness of a group is simply the summed fitness of its members; in the cooperative scheme the group fitness is determined by its fittest member. In other words, in the cooperative scheme, if at least one member in a group is fit, the entire group is protected from culling, reflecting the idea that the benefits of a key adaptive trait can be shared among all group members.

The details of how this is implemented is given in the next section and in Sec.~\ref{sec:results} the results are presented, showing that cooperation, in this example, does lead to a better evolutionary outcome. Although there are many limitations to this study, and some of these are discussed in Sect.~\ref{sec:discussion}, the way that the model demonstrates a Baldwin Effect for Cooperation in a robust way, without a broad parameter search, supports the possibility that this effect is important, for example, in the evolution of social traits.

\section{Methods}\label{sec:methods}

To fix terminology, each agent corresponds to a distinct rule for the cellular automaton and `rule' and `agent' are treated as being almost synonymous. A world is a one-dimensional grid of states that the rule is applied, potentially changing the states from iteration to iteration. In describing the simulations it is important to distinguish between evolution: the change from generation to generation of the collections of rules, and the fitness of agents, the number of successive application of the rule to the world before the world returns to empty.

A rule is evaluated by applying it successively starting with an initial world with just one cell in the state 1. A rule is considered to have stopped if it reaches a world with only empty cells. The challenge is deciding if a rule will ever stop and here an approximation has been used: a rule is regarded as going on for ever if a world state ever repeats since this would lead to periodic behaviour, or if an empty world has not resulted from 3000 applications of the rule; ruling out periodic behaviours is not sufficient, there are many rules that survive forever by growing larger and larger. The 3000 cutoff is chosen to keep the simulations computationally tractable, it does penalize the very fittest rules, but these are vanishingly rare. A rule that does not stop by the cutoff is given a fitness of -1, it is considered the least fit possible rule.
 
In all cases 100 rules are used in a simulation, arranged into 25 groups of four. Other arrangements, for example with groups of three or five, appear to produce very similar results and far larger numbers of trials would be need to decide what the optimal value of the group size is. At each generation the least fit group is culled and the most fit group is used to replace it; in the case of a draw, for either least or most fit, random selection is used. For the breeding scheme the fitness of a group is the sum of the fitness values for all four group members, for the cooperation scheme it is the maximum fitness. To produce the new group pairs of `parent' rules are picked from the fit group and a new rule is produced by selecting randomly from each parent for each clause with no bias towards more or less fit parent agents. 

The new rules produced during respawning are subject to mutation with probability $\lambda=0.75$; if a rule is selected for mutation one clause is selected at random and the output state of that clause is chosen with equal probability between the three possibilities. It should be noted that when the new child agent is mutated a single clause is chosen at random and its output state is selected randomly this will return the original rule in one case in three, so the mutation rate is effectively $2\lambda/3$. 

A scheme without breeding used as a comparison. In this case the group size is one but the population still has 100 rules. The only changes happen because of mutation. The mutation rate of $\lambda=0.75$ is used in all schemes: although a careful parameter search was not performed so parameter exploration was used and this value is chosen as a compromise between the lower mutation rate that would work best in breeding schemes and the waste of iterating the scheme without breeding without mutating the new rule. Four times as many generations are used in this scheme, since only one rule is culled at each generation.

\section{Results}\label{sec:results}

\begin{figure}[t]
\begin{center}
\begin{tabular}{lll}
&\textbf{A - breeding}&\textbf{B - cooperation}\\
\rotatebox{90}{\qquad \qquad \qquad fitness}
&\includegraphics[width=0.45\textwidth]{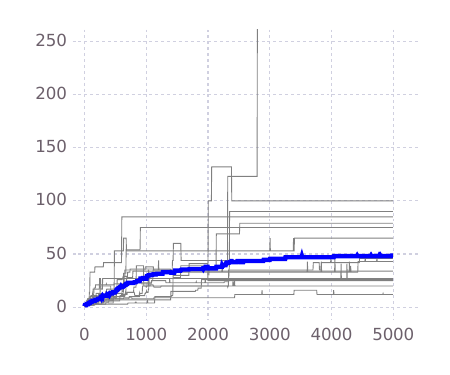}
&\includegraphics[width=0.45\textwidth]{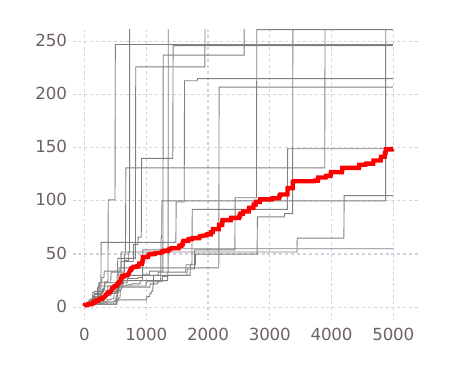}\\
&\qquad \qquad \qquad \qquad generations&\qquad \qquad \qquad \qquad generation
\end{tabular}
\end{center}
\caption{\textbf{Evolution in the model}. Two different evolutionary schemes have been run for a population of 100 agents for 5000 generations. Generation number is something of a misnomer, the 100 agents are divided into groups of four and at each generation the least-fit group is removed and replaced by descendant of the fittest group, so the lifespan of a typical agent is far greater than one generation, the average survival is 25 generations though, of course, survival will be very variable. In the \textbf{A - breeding} case the fitness of a group is the sum of the fitness of its individual member rules, in \textbf{B - cooperation} the fitness of a group is the fitness of the fittest member. In each case the scheme has been run for 100 trials. For the graph the fitness at each generation for a given trial is the fitness of the fittest rule. The red line is the median across trials, since evolution throws up the occasional rule whose fitness is considerable greater than the normal range of values, the median is used rather than the mean. The gray lines represent individual trials, although there were 100 trials for visual clarity a random subset of 15 are plotted. Again, for visual clarity, a fitness cutoff of 200 is used in the graphs, with some agents far exceeding this, the fittest agent in the collection that evolved in cooperation scheme has a fitness of 2792. \label{fig:evolution}}
\end{figure}

The main result of this paper is given in Fig.~\ref{fig:evolution} where the evolutionary progress of the system is plotted for two different schemes, one without cooperation, Fig.~\ref{fig:evolution}\textbf{A} and one with Fig.~\ref{fig:evolution}\textbf{B}. This simple version of cooperation is one where the fitness of group depends on the fittest member, rather than the sum of the fitness of all members. Here, the group size is four, with the whole population comprising of 25 groups. They are ordered by fitness, at each generation the least fit group is culled and respawned from the fittest group using a mixture of breeding and mutation, see Sect.~\ref{sec:methods} above for details. 

Cooperation gives a clear advantage, the median fitness across trials rises to 149 in the cooperation case, while for breeding it reaches 48.5. 
As a background comparison a scheme without breeding, that is with a group size of one and, obviously, no cooperation.
This scheme serves as a baseline: it retains the same population size but removes the potential for mixing rules, thereby illustrating how much of the performance benefit can be attributed to breeding or cooperation mechanisms.
The scheme without breeding reaches a median of 40. These numbers do in some ways undersell the advantage of cooperation, fitness becomes rarer and rarer as the value increases and . Let $c(f)$ be the number of agents with fitness greater than $f$, this can be estimated from a collection of randomly selected agents: here a collection of ten million random examples is used and gives $c(40)\approx 0.0014$ with:
\begin{equation}
    \frac{c(40)}{c(48.5)}\approx 2.20
\end{equation}
whereas 
\begin{equation}
    \frac{c(40)}{c(149)}\approx 29.2
\end{equation}
or $c(48.5)/c(149)\approx 13.3$ showing how much rarer an agent whose fitness matches the median in the cooperation scheme is compared to one whose fitness matches the media for the breeding scheme and that the sort of extremely fit rule that is discovered by the cooperation scheme is vanishingly rare. This demonstrates, albeit in an \textsl{ad hoc} way, just how much further the cooperative scheme has progressed in exploring the fitness landscape. 

It appears from Fig.~\ref{fig:evolution}\textbf{B} that the median fitness for the cooperation scheme has not finished increasing, in fact, the simulations were run for 12500 generations for 18 trials, the result has a median fitness of 276.5; at this point the value appears to have reached equilibrium. Performing longer runs in the breeding case indicates it had already reached equilibrium after the 5000 generations plotted in Fig.~\ref{fig:evolution}\textbf{A}.

\begin{figure}[tp]
\begin{center}
\begin{tabular}{ll}
&\textbf{A - breeding}\\
\rotatebox{90}{\qquad \qquad fitness}
&\includegraphics[width=0.95\textwidth]{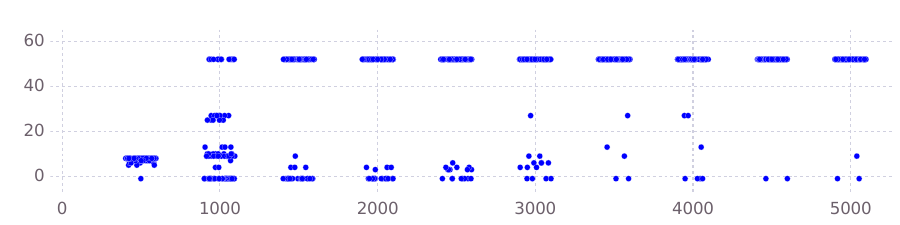}\\
&\textbf{B - cooperation}\\
\rotatebox{90}{\qquad \qquad fitness}&\includegraphics[width=0.95\textwidth]{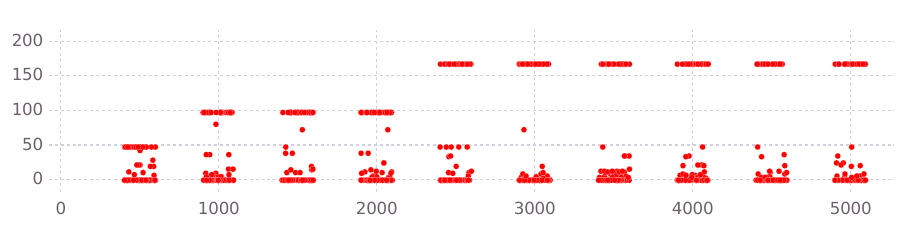}\\
&\textbf{C - number of unique rules}\\
\rotatebox{90}{\qquad \qquad unique}&\includegraphics[width=0.95\textwidth]{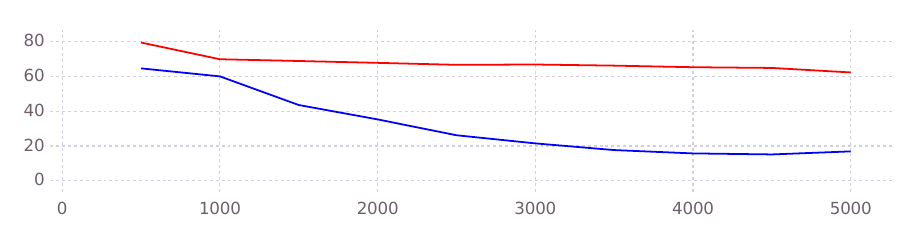}\\
&\qquad \qquad \qquad \qquad\qquad \qquad \qquad \qquad generation
\end{tabular}
\end{center}
\caption{\textbf{A typical trial}. Example trials are illustrated for the breeding scheme  (\textbf{A - breeding}) and the cooperation scheme (\textbf{B - cooperation}). Starting at 500, the fitness values of all 100 rules is plotted every 500 generations; to make the individual points more visible a small amount of horizontal jitter. Note that the horizontal axis is different in each case and in \textbf{A} the final fitness value is 52, whereas in \textbf{B} it is 167. In each case the trial was chosen to have final fitness values close to the median for each of the two schemes, this meant running the simulation for 12 trials for each scheme and selecting the one with the most suitable final fitness value. The same behaviour was visible in all 12 trials for each scheme. In \textbf{C - number of unique rules}, the number of different rules among the 100 in the population is plotted against generation for the breeding scheme (blue) and cooperation scheme (red); the line represents the average of 12 trials.\label{fig:trial}}
\end{figure}

The expectation is that the cooperation scheme evolves greater fitness because it maintains a broader population of agents, with less fit rules, as it were, benefiting from the protection or support of their fitter group-mates. To examine this the diversity of fitness values found in the population is plotted in Fig.~\ref{fig:trial} for a single trial. It is very clear that in the breeding trial Fig.~\ref{fig:trial}\textbf{A} the population becomes heterogenous with 97 out of the 100 rules having fitness 52 at generation 5000. In contrast, in the cooperation trial Fig.~\ref{fig:trial}\textbf{B} there continues to be a diversity of rules, with only 29 having fitness 167 at generation 5000. The same conconclusion is illustrated in Fig.~\ref{fig:trial}\textbf{C} which plots the number of unique rules as a function of generation; this falls sharply for the breeding scheme but not the cooperation scheme.

\section{Discussion}\label{sec:discussion}

The cellular automaton test for evolutionary schemes proposed in \cite{Wolfram2025} provides an interesting arena for looking at how the dynamics of evolution are affected by the choice of which agent is replaced and how it is respawned. It has an extremely complex fitness landscape. This system provides an illustration of a Baldwin Effect for Cooperation in which cooperation broadens the evolutionary path of a population by allowing less-fit members to survive and breed, retaining survivalist evolutionary dynamics while maintaining a diversity in the population which aids evolution.

Here only the three-state automaton has been considered, this has a huge number of possible rules. There are simpler examples, the two-model obviously, but it is also noted in \cite{Wolfram2025} that the three-state model with a symmetry constraint provides an example midway in simplicity between the two-state automaton and the full three-state automaton. In the future these simpler schemes could be used to describe the evolutionary dynamics in simulations of the Baldwin Effect for Cooperation in a more detailed way.

A larger variety of evolutionary schemes could be considered, for example, here there is no inter-group breeding and there has been no effort to search across parameter values even for the schemes that are considered. The intention here was to check the Baldwin Effect for Cooperation could be observed in a simulation based test of evolution without any ``finetuning'' of that test. It is also important to recognize that finding optimal parameter values might not provide much insight into evolution, there is no straight-forward mapping between aspects of this simulated evolution and biological evolution. It would, however, be interesting to test the stability of the results reported here as parameter values and other details of the scheme are changed and to study the evolution dynamics of the Baldwin Effect in more detail.

The evolutionary dynamics that produce complex characteristics can seem counter intuitive, particularly when partial realization of the characteristic does not appear to evolutionary benefit, the famous `what use is half an eye' problem. This is particularly interesting when the complex characteristics are behavioural or cognitive. One crucial example is the evolution of language where agent-based simulations have some potential to provide insight \cite{Kirby2000,KirbyHurford2002,Kirby2002a,BunyanBullockHoughton2024,BullockHoughton2024}. In this context, it is interesting to observe that there is a potential interaction between cooperative behaviour and evolution.

\subsection*{Code availability}
All the code and the data used to produce the figures is available \footnote{\texttt{github.com/BaldwinEffect/2025$\_$cellular}}. Julia 1.10.3 was used and run on Blue Crystal at the University of Bristol.
%
%

\end{document}